\documentclass[aps,prd,preprint,tightenlines,superscriptaddress,nofootinbib]{revtex4-1}
\usepackage{epsfig}
\usepackage{graphicx}
\usepackage{psfrag}
\usepackage{amsmath,amssymb}
\usepackage{colordvi}
\usepackage{color}
\usepackage{amsfonts}
\usepackage{enumerate}
\usepackage{slashed,bm}

\newcommand{\beq}{\begin{eqnarray}}
\newcommand{\eeq}{\end{eqnarray}}

\begin{document}

\begin{flushright}
YITP-13-111
\end{flushright}

\title{Gluon Helicity $\Delta G$ from a Universality Class \\ of Operators on a Lattice}
\author{Yoshitaka Hatta}
\affiliation{Yukawa Institute for Theoretical Physics, Kyoto University, Kyoto 606-8502, Japan}
\author{Xiangdong Ji}
\affiliation{INPAC, Department of Physics and Astronomy, Shanghai Jiao Tong University, Shanghai, 200240, P. R. China}
\affiliation{Center for High-Energy Physics, Peking University, Beijing, 100080, P. R. China}
\affiliation{Maryland Center for Fundamental Physics, University of Maryland, College Park, Maryland 20742, USA}

\author{Yong Zhao}
\affiliation{Maryland Center for Fundamental Physics, University of Maryland, College Park, Maryland 20742, USA}

\date{\today}
\vspace{0.5in}
\begin{abstract}
We show that the total gluon helicity $\Delta G$ in a polarized nucleon can be calculated on a Euclidean lattice through
a universality class of QCD operators that describe the helicity or polarization of the onshell gluon radiation.
We in particular find some operators whose matrix elements in a nucleon of momentum $P^z$ are directly related to $\Delta G$ with only power-law $(1/P^z)^n~ (n\ge 2)$ corrections.

\end{abstract}

\maketitle

\section{Introduction}

The total gluon helicity $\Delta G= \int^1_0 \Delta g(x)dx $ in a longitudinally polarized nucleon (proton or neutron) is an important physical
quantity that characterizes the fundamental property of the nucleon. In the last two decades, many high--energy experiments
have been carried out to measure the polarized gluon parton helicity distribution $\Delta g(x)$, from which one can
estimate the total gluon polarization by integration over the measured region, $\int^{x_{\rm max}}_{x_{\rm min}} dx \Delta g(x)$~\cite{Compass,Hermes,Phenix,Star}. Since
$\Delta G$ is intrinsically related to the light--cone physics, it has been impossible to calculate this quantity on a Euclidean lattice. Thus,
there have been little interplay between experiment and quantum chromodynamics (QCD) in this field so far~\cite{Jaffe:1995an,Chen:2006ng}.

In a recent publication~\cite{Ji:2013fga}, a theoretical method has been proposed to allow computing $\Delta G$ directly on a lattice
for the first time. Instead of the light--cone operator, the matrix element of a time--independent spin operator $\tilde \Delta G(P^z, \mu)$
is calculated in a nucleon with finite momentum $P^z$. The physical quantity $\Delta G$ is then obtained
through a matching condition
\begin{equation}
 \Delta \tilde G(P^z, \mu) = Z_{gg}(P^z/\mu) \Delta G(\mu) + Z_{gq}(P^z/\mu) \Delta \Sigma(\mu)\ ,
\label{matching}
\end{equation}
where $\Delta \Sigma(\mu)$ is the quark spin, and $\mu$ is the renormalization scale. $Z_{gg}$ and $Z_{qg}$ are the matching
coefficients calculable in QCD perturbation theory. The operator considered in Ref.~\cite{Ji:2013fga} was $\vec{E}\times \vec{A}_\perp$, where $\vec{A}_\perp$ is the transverse part of the gauge field, or $\vec{E}\times \vec{A}$ in the Coulomb gauge.

In this paper, we show that the gluon spin operator that can be matched to $\Delta G$ is not unique. Instead, one can find a
universality class of operators which can fulfill the same role. The physics of this phenomenon is easy to understand:
According to the Weizs\"acker--Williams approximation~\cite{Jackson}, the gluon field in the nucleon is dominated by quasi-free radiation which corresponds
to a beam of free gluons with momentum $\vec{k}=(0,0,xP^z)$. For such radiation, the gluon polarization vector is just
$\epsilon^\mu= (0,\epsilon^x, \epsilon^y,0)$. Thus, the gauge--dependent gluon spin operator $(\vec{E}\times \vec{A})^z=E^xA^y-E^yA^x$
under any gauge choice without changing the transverse polarization can describe the gluon helicity. These operators define a
universality class. For instance, in the Coulomb gauge, the gauge condition $\vec{k}\cdot \vec{A}=0$ yields $\epsilon^z=0$ which has no effect
on the spin operator. Another reason for the existence of a universality class is that the $t$--component and $z$--component of a four vector scale in the same way in the infinite momentum frame (IMF) limit.

In Sec.~II, we explore different gluon spin operators that correspond to different gauge choices for
$\vec{E}\times \vec{A}$, and show that they all lead to the same light--cone gluon helicity $\Delta G$. We consider physical gauges as well as covariant
gauges. In Sec.~III, we consider the matrix element of the topological current, leading to some more operators of the universality class
which do not even have straightforward gluon spin interpretation. We consider their matrix elements to one--loop in the continuum in order to provide a useful input for matching to lattice QCD calculations. We conclude the paper in Sec.~IV.

\section{A Universality Class of Operators}

In this section, we discuss the matrix elements of the gluon spin operator with different choices of gauges, which asymptotically
approach the physical gluon helicity $\Delta G$. We start with the consideration in Ref.~\cite{Ji:2013fga}.

Let us begin with the standard definition of $\Delta G$ as the matrix element of a
non-local operator involving light--cone correlation~\cite{Manohar:1990jx}
\begin{eqnarray}
  \Delta G \frac{S^+}{P^+} &=& \int dx \frac{i}{2xP^+} \int
  \frac{d\xi^-}{2\pi} e^{-ixP^+\xi^-} \langle PS| F^{+\alpha}_a(\xi^-) {\cal L}^{ab}(\xi^-,0)\tilde F_{\alpha,b}^{~+} (0)|PS\rangle_N \nonumber \\
 &=& \frac{1}{2P^+}\langle PS| \epsilon^{ij}F^{i+}(0)A^j_{\rm phys}(0)|PS\rangle_N\,,
\label{gpdf}
\end{eqnarray}
where  $|PS\rangle_N$ is a proton plane--wave state with momentum $P^\mu$ and polarization $S^\mu$, $\tilde F^{\alpha\beta} = (1/2)\  \epsilon^{\alpha\beta\mu\nu}F_{\mu\nu}$,
and ${\cal L}(\xi^- ,0) = P\exp[-ig\int^{\xi^-}_0  A^+(\eta^-,0_\perp)\ d\eta^-]$
is a  gauge link  in the adjoint representation. The light--front coordinates are defined as $\xi^\pm = (\xi^t\pm \xi^z)/\sqrt{2}$.

In the second line of Eq.~(\ref{gpdf}), we defined~\cite{Hatta:2011zs,Leader:2013jra}
  \beq
A^\mu_{\rm phys}\equiv\frac{1}{D^+}F^{+\mu}\,, \label{def}
  \eeq
and introduced the antisymmetric tensor in the transverse plane
  $\epsilon^{ij}$ ($\epsilon^{xy}=-\epsilon^{yx}=1$). The boundary condition for the integral operator $1/D^+$ is related to the $i\epsilon$--prescription for the $1/x$ pole.
 In the light--cone gauge $A^+=0$, $A^\mu_{\rm phys}$ reduces to $A^\mu$.

The matrix element in Eq.~(\ref{gpdf}), being nonlocal in the light--cone direction, cannot be readily evaluated in lattice QCD. However, it has been suggested in Ref.~\cite{Ji:2013fga} that one can relate $\Delta G$ to the following matrix element
 \beq
\Delta \tilde G(P^z, \mu) =  \frac{1}{2P^0}\langle PS| \epsilon^{ij}F^{i0}(0)A^j(0)|PS\rangle_N\,, \label{gauge}
\eeq
 which is local, hence measurable on the lattice. In Eq.~(\ref{gauge}), the momentum $P^z$ is assumed to be large but finite.
  $\epsilon^{ij}F^{i0}A^j=(\vec{E}\times \vec{A})^z$  is the gluon helicity operator identified by Jaffe and Manohar \cite{Jaffe:1989jz}. As is well--known, this operator is not gauge invariant, so the matrix element in Eq.~(\ref{gauge}) depends on the gauge choice.  In Ref.~\cite{Ji:2013fga} the authors used the Coulomb gauge (see Refs.~\cite{Chen:2008ag,Chen:2009mr} for an earlier discussion)
 \beq
\vec{\nabla}\cdot \vec{A}=0\,. \label{trans}
  \eeq
The condition in Eq.~(\ref{trans}) separates the transverse (or ``physical") part from the gauge field which should be kept in the computation of physical quantities like $\Delta G$.  While the solution $A^\mu=A^\mu_\perp$ to Eq.~(\ref{trans}) in generic frames bears no resemblance to $A^\mu_{\rm phys}$, it has been shown in Ref.~\cite{Ji:2013fga} that $A^\mu_\perp$ approaches $A^\mu_{\rm phys}$ if one takes the IMF limit.\footnote{See, Eq.~(7) of Ref.~\cite{Ji:2013fga} and notice that in the IMF limit,
  \beq
  A_\perp^\mu \to A^\mu -\frac{1}{D^+}\partial^\mu A^+ = \frac{1}{D^+}F^{+\mu} = A_{\rm phys}^\mu\,.
  \eeq
  }

In field theory, due to ultraviolet (UV) divergences, a subtlety arises that the matrix elements involving $A_{\rm phys}^\mu$ and $A_\perp^\mu$
are not simply related by a Lorentz boost. For the external onshell quark state $|PS\rangle_q$, the one--loop  calculation using dimensional regularization (in $D=4-2\epsilon$ dimensions) yields~\cite{Chen:2011gn,Ji:2013fga}
  \beq
  \Delta \tilde G(P^z, \mu)  = \left.\frac{\langle PS| \epsilon^{ij}F^{i0}A^j|PS\rangle_q}{2P^0} \right\arrowvert_{\vec{\nabla}\cdot \vec{A}=0} =
\frac{C_F\alpha_s}{4\pi} \left(\frac{5}{3\varepsilon_m}-\frac{1}{9}+ \frac{4}{3}\ln \frac{4P_z^2}{m^2}\right)\frac{S^z}{P^0} \, ,  \label{p}
\eeq
 where we defined $1/\varepsilon_m \equiv 1/\epsilon-\gamma_E + \ln 4\pi + \ln (\mu^2/m^2)$, and $m$ is the quark mass to regularize the collinear divergence. $C_F=(N_c^2-1)/2N_c$ as usual. On the other hand, in the same regularization scheme Eq.~(\ref{gpdf}) is evaluated as
\beq
\Delta G(\mu) = \left. \frac{\langle PS|\epsilon^{ij}F^{i+}A^j|PS\rangle_q}{2P^+}\right\arrowvert_{A^+=0}=\frac{C_F\alpha_s}{4\pi}\left(
\frac{3}{\varepsilon_m}+7\right) \frac{S^+}{P^+}\,. \label{lc}
\eeq
 We see that the coefficients of $1/\varepsilon_m$ (anomalous dimension) are different. Moreover,  Eq.~(\ref{p}) depends nontrivially on the reference frame. The reason for this discrepancy is that the IMF limit $P^z\to \infty$ and the large loop momentum limit  $k^\mu\to \infty$ in the one--loop integral do not commute: One can actually recover the light--cone gauge result in Eq.~(\ref{lc}) from the Coulomb gauge calculation by sending $P^z\to \infty$ \emph{before} doing the $k$--integral. On a lattice, $P^z$ is restricted to be less than the cutoff, which is tantamount to taking the $k^\mu\to \infty$ limit first. Thus, the matrix element in Eq.~(\ref{gauge}), evaluated in the Coulomb gauge, fails to capture the UV properties of $\Delta G$.  Nevertheless, since the infrared (IR) physics is not affected by the order of limits, one can correct the discrepancy via the \emph{matching condition}
 \cite{Ji:2013fga}\footnote{On the lattice, $1/\varepsilon_m$ is replaced by $-\ln (a^2m^2)$ so the matching condition becomes $\ln P_z^2a^2=const.$}
\beq
\frac{1}{\varepsilon_m} + \frac{16}{3} =\ln\frac{4P_z^2}{m^2}\,. \label{coulomb}
\eeq
This observation paves the way for evaluating $\Delta G$ on a Euclidean lattice.

To see the relevance for nonperturbative calculations, we use the general matching formula as given by Eq.~(\ref{matching}).
According to the result from Eq.~(\ref{p}), we find, for the Coulomb gauge spin operator,
\begin{equation}
        Z_{gq}(P^z/\mu) = \frac{C_F\alpha_s}{4\pi}\left( \frac{4}{3}\ln \frac{4(P^z)^2}{\mu^2} - \frac{64}{9}\right)\,,
\end{equation}
in the $\overline{\mbox{MS}}$ scheme, which is IR-free. The matching coefficient $Z_{gg}$ must be calculated in a gluon state.\\

The Coulomb gauge in Eq.~(\ref{trans}) is not the unique possibility in order to match with $\Delta G$.
For instance, consider the temporal axial gauge $A^0=0$. In this gauge one can identically write
\beq
\tilde  A^\mu = \frac{1}{D^0}F^{0\mu}\,. \label{0}
 \eeq
 Taking the IMF limit, one trivially recovers Eq.~(\ref{def}),
 \beq
 \frac{1}{D^0}F^{0\mu} \to \frac{1}{D^+}F^{+\mu}=A_{\rm phys}^\mu\,.
 \eeq
Alternatively, one may choose the $A^z=0$ gauge in which
\beq
\tilde A^\mu = \frac{1}{D^z}F^{z\mu}\,. \label{z}
 \eeq
 This also becomes $\frac{1}{D^+}F^{+\mu}$ in the IMF limit.

However, as  in the Coulomb gauge, the matrix elements of the operator $\vec{E}\times\vec{A}$ are in general different. To one--loop order, we find
\beq
&&   \Delta \tilde G(P^z, \mu)  = \left.\frac{\langle PS|\epsilon^{ij}F^{i0}A^j|PS\rangle_q}{2P^0} \right\arrowvert_{A^0=0}=
\frac{C_F\alpha_s}{4\pi}\left(\frac{3}{\varepsilon_m}+7\right) \frac{S^z}{P^0}\,, \label{00} \\
&&  \Delta \tilde G(P^z, \mu)  =  \left.\frac{\langle PS|\epsilon^{ij}F^{i0}A^j|PS\rangle_q}{2P^0} \right\arrowvert_{A^z=0}=  \frac{C_F\alpha_s}{4\pi} \left[\frac{2}{\varepsilon_m}+4+ \frac{P^z}{P^0}\ln \frac{(P^0+P^z)^2}{m^2}\right]\frac{S^z}{P^0}\,. \label{zz}
\eeq
Eq.~(\ref{00}) agrees with the previous result in Eq.~(\ref{lc}) in the light--cone gauge (see, also, Ref.~\cite{Wakamatsu:2013voa}). On the other hand, Eq.~(\ref{zz}) features yet another anomalous dimension together with logarithmic frame dependence. Here again, the order of limits matters: If one takes the $P^z\to \infty$ limit before the loop integration, one recovers Eq.~(\ref{lc}) from the $A^z=0$ gauge calculation. At large but finite momentum, part of the divergence $1/\varepsilon_m$ is transferred to the logarithm $\ln P_z^2$, keeping the sum of their coefficients unchanged. The following matching condition then establishes the connection between Eq.~(\ref{zz}) and Eq.~(\ref{lc}),
  \beq
\frac{1}{\varepsilon_m}+3 =\frac{P^z}{P^0}\ln \frac{(P^0+P^z)^2}{m^2} \approx \ln \frac{4P_z^2}{m^2}\,. \label{mat}
\eeq
The constant term is different from the Coulomb gauge case in Eq.~(\ref{coulomb}). This corresponds to a different matching constant $Z_{qg} = (C_F\alpha_s/4\pi)(\ln4(P^z)^2/\mu^2 - 3)$.

Thus, for the purpose of obtaining $\Delta G$, one can broadly generalize the approach of  Ref.~\cite{Ji:2013fga}: Evaluate the ``naive" gluon helicity operator Eq.~(\ref{gauge}) either in the Coulomb gauge, or $A^0=0$, or $A^z=0$ gauge and perform an appropriate matching. However, this does not mean that any gauge choice is allowed.
For instance, in the $A^x=0$ gauge where
\beq
\tilde A^\mu = \frac{1}{D^x}F^{x\mu}\,, \label{ax}
 \eeq
  or in the Landau (or covariant) gauge $\partial \cdot A=0$ where
 \beq
\tilde A^\mu=A^\mu -\frac{1}{\partial^2}\partial^\mu \partial \cdot A\,,
 \eeq
  $\tilde A^\mu$ does not approach $A_{\rm phys}^\mu$ in the IMF limit. This is also reflected in their one--loop matrix elements
 \beq
&& \left.\frac{\langle PS|\epsilon^{ij}F^{i0}A^j|PS\rangle_q}{2P^0} \right\arrowvert_{A^x=0}=
\frac{C_F\alpha_s}{4\pi}\left(\frac{3}{2\varepsilon_m}+\frac{7}{2}\right) \frac{S^z}{P^0}\,,  \label{x0} \\
&& \left.\frac{\langle PS|\epsilon^{ij}F^{i0}A^j|PS\rangle_q}{2P^0} \right\arrowvert_{\partial  \cdot A=0}=  \frac{C_F\alpha_s}{4\pi} \left(\frac{2}{\varepsilon_m}+4\right)\frac{S^z}{P^0}\,,
\eeq
 which do not agree with the light--cone gauge result.\footnote{Interestingly,  Eq.~(\ref{x0}) is exactly one half of Eq.~(\ref{00}).} Moreover, the logarithm of $P^z$ is absent so there is no possibility of matching.\\

The above analysis suggests that there is a class of gauges (similar to the universality class of second order phase transitions) which flows to the ``fixed point" $A_{\rm phys}$ in the IMF limit, and thus can be used to compute $\Delta G$. This class of gauges clearly do not include all possible gauges. 
To see what gauges are permitted, we consider the Weizs\"acker--Williams (WW) approximation~\cite{Jackson} in the IMF. 
The gluon field is dominated by quasi-free radiation in the sense that $\vec{B}_\perp \sim \vec{E}_\perp \gg \vec{E}_{||}$.
Thus we have in effect a beam of gluons with momentum $xP^z$. For these onshell gluons, the gauge transformation only affects 
the time component and the third spatial component (we consider only the Abelian
part),
\begin{equation}
                     A^\mu \rightarrow A^\mu + \lambda k^\mu\,,
\label{gauge2}
\end{equation}
where $k^\mu = (k^0, 0, 0, k^z)$. Thus the transverse part of the polarization vectors is physical,
\begin{equation}
                      \epsilon^\mu(xP^z) = \frac{1}{\sqrt{2}}(0, 1, \mp i, 0) \ .
\end{equation}
The gluon spin operator $(\vec{E}\times \vec{A})^z$ is independent of those gauge transformations which leave
$A^{x,y}$ invariant. Although Eq. (\ref{gauge2}) seems to guarantee this for WW gluon field, it contains only a subclass of gauges: 
There are gauge choices which are incompatible with the notion that WW gluon $A^{x,y}$ shall be left intact by 
gauge transformations. Those latter gauge 
transformations will not ``flow" into the fixed point light--cone operator in the IMF. 

The axial gauge $A^z=0$ and the temporal gauge $A^0=0$ have no effect on the gluon polarization vector. Therefore,
they can be used to calculate the gluon helicity.
In the Coulomb gauge, one has  $\vec{k}\cdot \vec{A}=k^zA^z=0$. This is similar to the axial gauge $A^z=0$.

The obvious counterexample is $A^x=0$ or $A^y=0$ gauges. A less trivial one is the covariant gauge, in which 
the condition $k\cdot A = k^+A^-=0$ itself is consistent with having nonzero transverse components. However, actually the WW field in the covariant gauge  has only the $A^+$ component. This can be seen from an example of the WW field associated with a fast--moving  pointlike charge. In the covariant gauge we have
\beq
A^\mu(\xi) =  -e\ln \xi_\perp^2 \delta(\xi^-) \delta^\mu_+\,. \label{ww}
\eeq
Eq.~(\ref{ww}) indeed satisfies $\partial \cdot A=\partial_+A^+=0$, but has vanishing transverse components $A^{x,y}$. Therefore the covariant gauge does not belong to the universality class.



\section{Axial gauges, topological current, and more operators}

The temporal axial gauge $A^0=0$ seems to have a special status since the matrix element in Eq.~(\ref{00}) coincides with that in the $A^+=0$ gauge. Therefore, in this section we explore strategies to measure $\Delta G$ in the $A^0=0$ gauge where there is no logarithmic matching, or more generally, in non-lightlike axial gauges $n\cdot A=0$ with $n^2\neq 0$ (see, also, Ref.~\cite{Wakamatsu:2013voa}). As we shall see,
the matrix element of the topological current allows us to find more operators in the universality class, and some of them do not even have the form of spin operator in a particular gauge.

First, note that
in the $A^0=0$ gauge, the operator $\epsilon^{ij}F^{i0}A^j$ is the same as
\beq
\epsilon^{ij} \left(F^{i0}A^j-\frac{1}{2}A^0F^{ij}\right)\,. \label{mod}
\eeq
Likewise, in the $A^+=0$ gauge the operator $\epsilon^{ij}F^{i+}A^j$ is the same as
\beq
\epsilon^{ij} \left(F^{i+}A^j-\frac{1}{2}A^+F^{ij}\right)\,. \label{mod2}
\eeq
Actually, the matrix elements of these operators are gauge invariant to one--loop,
\beq
\frac{\langle PS|\epsilon^{ij}\left(F^{i0}A^j-\frac{1}{2}A^0F^{ij}\right)|PS\rangle_q}{2P^0} =\frac{C_F\alpha_s}{4\pi}\left(\frac{3}{\varepsilon_m}+7\right) \frac{S^z}{P^0}\,, \label{now}
\eeq
as can be explicitly checked in all the gauges mentioned in the previous section. [See, also,   Ref.~\cite{Guo:2012wv}.]
 This in particular means that the logarithm of $P^z$ which appears in some gauges is canceled by the contribution from the extra term $\epsilon^{ij}A^0F^{ij}$.
The reason is that Eqs.~(\ref{mod}) and (\ref{mod2}) are a part of the topological current in QCD
\beq
 K^\mu &=&\epsilon^{\mu\nu\rho\lambda} \left(A_\nu^a F_{\rho\lambda}^a + \frac{g}{3}f_{abc}A_\nu^a A_\rho^b A_\lambda^c \right) \,, \label{top} \\
 K^+&=& 2\epsilon^{ij} \left(F_a^{i+}A_a^j - \frac{1}{2}F_a^{ij}A_a^+ - \frac{g}{2}f_{abc}A^+_a A^b_i A^c_j\right)\,, \nonumber \\
 K^z&=& 2\epsilon^{ij}\left(F_a^{i0}A_a^j - \frac{1}{2}F_a^{ij}A_a^0 - \frac{g}{2}f_{abc}A^0_a A^b_i A^c_j \right)\,, \nonumber
 \eeq
 which satisfies $\partial_\mu K^\mu=F^{\mu\nu}_a\tilde{F}^a_{\mu\nu}$. The forward matrix element of Eq.~(\ref{top}) is perturbatively gauge invariant \cite{Altarelli:1988nr,Jaffe:1989jz} and the ${\mathcal O}(gAAA)$ term starts to contribute only at two loops (for quark external states).

Nonperturbatively, however, there is gauge dependence due to anomaly~\cite{Jaffe:1989jz,Manohar:1990eu,Balitsky:1991te}.
In axial gauges $A\cdot n=0$, this dependence has been precisely calculated in Ref.~\cite{Balitsky:1991te}. The \emph{non}-forward matrix element of $K^\mu$ in a polarized nucleon state is given by
\begin{equation}
 \langle PS|K^\mu |P+q,S\rangle_N\Big\arrowvert_{A\cdot n=0} \xrightarrow{q^\mu\to 0} 4\left(S^\mu -\frac{q\cdot S}{q\cdot n} n^\mu \right)\Delta G\left(n,P\right) + \frac{in^\mu}{q\cdot n}\langle PS|F_a^{\mu\nu}\tilde{F}^a_{\mu\nu} |PS\rangle_N\,, \label{e}
\end{equation}
where
\beq
\int_0^\infty d\lambda \langle PS| n^\tau F_{\tau\nu}(\lambda n){\cal L}\tilde{F}^{\nu\mu}(0)|PS\rangle_N \equiv 2S^\mu \Delta G(n,P)\,. \label{ba}
\eeq
The matrix element in Eq.~(\ref{ba}) is the same as in Eq.~(\ref{gpdf}) except for the direction of the Wilson line. Expanding around the deviation from the light--cone $n^2$, one finds the relation \cite{Balitsky:1991te}
 \beq
\Delta G(n,P) =\ \Delta G + {\mathcal O}\left(\frac{n^2}{(P\cdot n)^2}\right)\,, \label{braun}
\eeq
which is valid at large momentum  (assuming $P\cdot n\neq 0$).

From Eq.~(\ref{e}) one can read off various representations of $\Delta G$. For the $\mu=z$ component in the $A^0=0$ gauge, the ambiguity (gauge dependence) in the $q^\mu\to 0$ limit drops out. One can safely take the forward limit and  find
\beq
\left.\langle PS|\epsilon^{ij}A^i \partial^0 A^j |PS\rangle_N \right\arrowvert_{A^0=0} = 2S^z\Delta G + {\mathcal O}(1/P_z^2)\,. \label{q1}
\eeq
This result extends Eq.~(\ref{00}) to all orders in perturbation theory.
Similarly, taking $\mu=0$ in the $A^z=0$ gauge, one gets
\beq
\left.\langle PS|\epsilon^{ij}A^i \partial^z A^j |PS\rangle_N \right\arrowvert_{A^z=0} = 2S^0\Delta G + {\mathcal O}(1/P_z^2)\,, \label{q2}
\eeq
which is related to Eq.~(\ref{zz}) by replacing $F^{i0}$ with $F^{iz}$. In the IMF limit, the $t$--component and $z$--component of a 
quantity have similar scaling properties as they both approach the plus ($+$) direction. 
Note that the operator on the left hand side of Eq.~(\ref{q2}) does
not have a straightforward gluon spin interpretation.

Moreover,  Eqs.~(\ref{ba}) and (\ref{braun}) directly give
\beq
\int_0^\infty d\xi^0 \langle PS|  F^0_{\ \nu}(\xi^0){\cal L}\tilde{F}^{\nu 0}(0)|PS\rangle_N &=& \left.\langle PS|\vec{A}^a\cdot \vec{B}^a |PS\rangle_N \right\arrowvert_{A^0=0} \nonumber \\
&=& 2S^0\Delta G + {\mathcal O}(1/P_z^2)\,. \label{p1}
\eeq
\beq
\int_0^\infty d\xi^z \langle PS|  F^{z}_{\ \nu}(\xi^z){\cal L}\tilde{F}^{\nu z}(0)|PS\rangle_N &=&\left.\langle PS|\epsilon^{ij}\left(F^{i0}A^j-\frac{1}{2}A^0F^{ij}\right)|PS\rangle_N \right\arrowvert_{A^z=0} \nonumber \\ &=& 2S^z\Delta G+ {\mathcal O}(1/P_z^2)\,. \label{x}
\eeq
The operator in Eq. (\ref{p1}) is similar to an operator written down by Jaffe~\cite{Jaffe:1995an}, except that it includes the $z$--component as well. 
Eq.~(\ref{x}) coincides with the operator introduced in Ref.~\cite{Ji:2013dva}.
 All the matrix elements in Eqs.~(\ref{q1})--(\ref{x}) are measurable on the lattice. In particular,  the operators in Eqs.~(\ref{q2}) and (\ref{p1}) can be readily transcribed into Euclidean space as they do not contain temporal indices  $\partial^0$, $A^0$. Note that all these operators yield the gluon helicity $\Delta G$ without logarithmic corrections in the large $P^z$ limit.

\subsection{Matching on Lattice}
In order to relate $\Delta G_{\rm lat}$ measured on the lattice to $\Delta G_{\rm \overline{MS}}$ defined in the continuum theory in the $\overline{\mbox{MS}}$ scheme, one has to perform a perturbative matching. The matching coefficients depend on the operators chosen and the UV regularization (independent of the IR regulator). In the case
of the operators discussed above, the perturbative matching is particularly simple because there are no large logarithms $\ln P^z/\mu$ involved.

We first consider the mixing of $\Delta G$ with the quark spin $\Delta \Sigma$. This can be read off from Eq.~(\ref{lc}), but here  we use a different regularization of the collinear divergence in order to keep in line with the gluon matrix element calculated below, and also with typical lattice computations \cite{Capitani:2002mp}. Namely, we now assume that the quark is massless and slightly off--shell $P^2<0$. This affects the finite term of the matrix element
\beq
 \left. \langle PS|\epsilon^{ij}F^{i+}A^j|PS\rangle_q \right\arrowvert_{A^+=0}&=&\frac{C_F\alpha_s}{4\pi}\left(
\frac{3}{\varepsilon_v}+4\right) \langle PS|\bar{q}\gamma_5\gamma^+ q|PS\rangle_q^{tree} \,,
 \label{new}
\eeq
 where $1/\varepsilon_v \equiv 1/\epsilon-\gamma + \ln 4\pi + \ln \frac{\mu^2}{-P^2}$. We remind the reader that the  nucleon matrix element of the quark operator on the right hand side is related to $\Delta \Sigma$ as $
 \langle PS|\bar{q}\gamma_5\gamma^\mu q|PS\rangle_N = 2S^\mu \Delta \Sigma$.
Due to the facts that the operator $K^\mu$ transforms as a Lorentz vector and is one--loop gauge invariant, Eq.~(\ref{new}) immediately implies that the same coefficient should appear in the (quark) matrix element of all the operators in Eqs.~(\ref{q1})--(\ref{x}), e.g.
\beq
&& \left. \frac{\langle PS|\epsilon^{ij}F^{i0} A^j |PS\rangle_q}{2S^z} \right\arrowvert_{A^0=0} = \left. \frac{\langle PS|\epsilon^{ij}A^i \partial^z A^j |PS\rangle_q}{2S^0} \right\arrowvert_{A^z=0} =\frac{\alpha_s C_F}{4\pi}\left(\frac{3}{\varepsilon_v}+4\right)\,. \label{waka}
\eeq

  Next we compute the one--loop matrix element in the gluon external state $|Ph\rangle_g$ ($h=\pm 1$ is the helicity).
 In the light--cone gauge with the Mandelstam--Leibbrandt prescription for the propagator pole $1/k^+\to 1/(k^+ +i\epsilon k^-)$, the contribution of the irreducible diagram is calculated to be (see Appendix \ref{gluon})
 \beq
\left.\frac{\langle Ph | \epsilon^{ij}F^{i+}A^j |Ph\rangle_g}{2P^+} \right|^{irr}_{A^+=0}  =h\frac{\alpha_s N_c}{2\pi} \left(2 + \frac{\pi^2}{3} \right)\,. \label{glue}
\eeq
 Note that there is no divergence.
The self--energy insertion in the external gluon legs is divergent and reads (cf. Ref.~\cite{Dalbosco:1986eb})
\beq
\left.\frac{\langle Ph | \epsilon^{ij}F^{i+}A^j |Ph\rangle_g}{2P^+} \right|^{self}_{A^+=0} = h \frac{\alpha_sN_c}{2\pi} \left(\frac{11}{6\varepsilon_v} -\frac{\pi^2}{3} + \frac{67}{18}\right) + h\frac{\alpha_s N_f}{2\pi} \left(-\frac{1}{3\varepsilon_v} -\frac{5}{9} \right)\,, \label{twoterm}
\eeq
  where the two terms correspond to the gluon and quark loop contributions, respectively.
 Combining these results,  we find
\beq
\left.\langle Ph | \epsilon^{ij}F^{i+}A^j |Ph\rangle_g \right|_{A^+=0}= \left[ 1+\frac{\alpha_s}{4\pi} \left( \frac{\beta_0}{\varepsilon_v} + \frac{103N_c-10N_f}{9}  \right) \right]\langle Ph|\epsilon^{ij}F^{i+}A^j |Ph\rangle_g^{tree} \,, \label{finite}
\eeq
where $\beta_0=\frac{11N_c}{3}-\frac{2N_f}{3}$ is the coefficient of the one--loop QCD beta function.
By the same reasoning as in Eq.~(\ref{waka}), we immediately obtain\footnote{The agreement of the divergent part in Eqs.~(\ref{finite}) and (\ref{wa2}) was  explicitly  checked in Ref.~\cite{Wakamatsu:2013voa}.}
\beq
\left.\langle Ph | \epsilon^{ij}F^{i0}A^j |Ph\rangle_g \right|_{A^0=0}= \left[ 1+\frac{\alpha_s}{4\pi} \left( \frac{\beta_0}{\varepsilon_v} + \frac{103N_c-10N_f}{9}  \right) \right]\langle Ph|\epsilon^{ij}F^{i0}A^j |Ph\rangle_g^{tree} \,, \label{wa2}
\eeq
 and similarly for the other matrix elements in Eqs.~(\ref{q2})--(\ref{x}). Note that, \emph{a priori}, the one--loop calculation of the latter two matrix elements Eqs.~(\ref{p1}) and (\ref{x}) could be complicated, not least because the non--Abelian part of the operator ${\mathcal O}(gAAA)$ would contribute already at one--loop for gluon external states. Yet, the above discussion guarantees that the final result is identical to the one computed in the light--cone gauge Eq.~(\ref{finite}).
 In the $\overline{\mbox{MS}}$ scheme, $1/\varepsilon_v$ is replaced by $\ln \mu^2/(-P^2)$. In lattice perturbation theory the logarithms become
 $\ln 1/(a^2P_E^2)$ and their coefficients (anomalous dimensions) are the same. The matching of the constant terms can be done in a standard manner \cite{Capitani:2002mp}.\footnote{We note that there exists an exact matching scheme \cite{Ji:1995vv} which goes beyond the one--loop matching considered here. }
\\
\section{conclusion}

In this paper, we first extended the matching method of Ref.~\cite{Ji:2013fga}  to a broader class of gauges. Not only the Coulomb gauge, but also other gauge choices that maintain the $A^{x,y}$ components of the onshell gluon fields do qualify, and in some of them the gluon spin matrix element does not have logarithmic corrections in the large momentum limit. We then focused our attention on non-lightlike axial gauges. All the matrix elements in Eqs.~(\ref{q1})--(\ref{x}) can be used to compute $\Delta G$ in lattice QCD, and we have computed the one--loop matching coefficients on the continuum theory side.

The implementation of the Coulomb gauge and  axial gauges on a lattice may pose technical problems. The usual periodic boundary condition on gauge field configurations is incompatible with the  condition $A\cdot n=0$ because of nonvanishing Polyakov loops. In order to circumvent this and fix the residual gauge symmetry, ideally one should impose antisymmetric boundary condition in the direction specified by the vector $n^\mu$. Or else, one has to confront the problem of lattice Gribov copies~\cite{Sharpe:1984vi,Neuberger:1986xz}.
\\

\vspace{2em}
We thank S.~Aoki and J.~-W. Qiu for useful discussions.
This work was partially supported by the U. S. Department of Energy via
Grant No. DE-FG02-93ER-40762, the Office of Science
and Technology of Shanghai Municipal Government Grant
No. 11DZ2260700, and by the National Science
Foundation of China Grant No. 11175114. Y. Hatta and X. Ji also thank 
RIKEN for their travel support, where this collaboration started.

\appendix

\section{Gluon matrix element of $(\vec{E}\times \vec{A})^z$}
\label{gluon}

In this appendix we display some intermediate steps leading to the result in Eq.~(\ref{glue}).
We treat the external gluon to be off--shell $P^2<0$. After some algebra, the one--loop matrix element in the light--cone gauge reduces to (see, also, Ref.~\cite{Wakamatsu:2013voa})
\begin{equation}
\frac{\langle Ph | \epsilon^{ij}F^{i+}A^j |Ph\rangle_g}{2P^+} \sim h \frac{ig^2N_c}{P^+}\int \frac{d^Dk}{(2\pi)^D} \frac{\frac{16}{D-2}k_\perp^2 P^+-k^+(P+k)^2 -2\frac{P^++k^+}{P^+-k^+}k^+(k^2-P^2)}{k^2k^2(P-k)^2} \,.
\end{equation}
 We use the Mandelstam--Leibbrandt prescription for the pole in the last term of the numerator
 $1/k^+ \to 1/(k^+ + i\epsilon k^-)$. The following formulas are useful:
 \beq
&& \int \frac{d^Dk}{(2\pi)^D} \frac{1}{k^2(P-k)^2 (P^+-k^+)} =  \frac{i}{(4\pi)^2 P^+} \frac{\pi^2}{6}\,,
\eeq
\beq
&&\int  \frac{d^Dk}{(2\pi)^D} \frac{1}{k^2k^2(P-k)^2 (P^+-k^+)} = \frac{-i}{(4\pi)^2P^+ P^2} \frac{1}{\varepsilon'_v}\,,
\eeq
where $\varepsilon'_v$ is an IR regulator.

\end{document}